\documentclass[useAMS,usenatbib]{mn2e}

\usepackage[dvips]{graphicx}
\usepackage{graphics}
\usepackage{psfig}
\usepackage{lscape}
\usepackage{amssymb}
\input{epsf}

\title[ Hide-and-seek in H0557-385]{Obscuring clouds playing hide-and-seek in the Active Nucleus H0557-385 }
\author[A. L. Longinotti et al.]{A. L. Longinotti$^{1}$\thanks{E-mail:
alonginotti@sciops.esa.int}, S.Bianchi$^{2}$, L. Ballo$^{1}$, I. de la Calle$^{1}$, M.Guainazzi$^{1}$ \\
$^{1}$ XMM-Newton Science Operation Centre, RSSD-ESA, ESAC, P.O. Box 78, 28691 Villanueva de la Ca{\~n}ada, Madrid, Spain\\
$^{2}$ Dipartimento di Fisica, Universit\`a degli Studi Roma Tre, Via della Vasca Navale 84, 00146, Roma, Italy\\
}

\begin{document}

\date{Accepted}

\pagerange{\pageref{firstpage}--\pageref{lastpage}} \pubyear{2008}

\maketitle

\label{firstpage}

\begin{abstract}
This paper reports on two {\it XMM-Newton}  observations of the Seyfert 1 Galaxy H0557-385 obtained in 2006, which show the source at an historical low flux state, more than a factor of $\sim$10 lower than a previous {\it XMM-Newton} look in 2002. 
The low flux spectrum presents a strong Fe K$\alpha$ line associated to a Compton reflection continuum. An additional spectral line around 6.6~keV is required to fit K$\alpha$ emission from Fe XXV. The spectral curvature below 6~keV implies obscuration by  neutral gas with a column density of $\sim$8$\times$10$^{23}$~cm$^{-2}$ partially covering the primary emission, which still contributes for a few percent of the soft X-ray  emission.
Absorption by ionised material on the line of sight is required to fit the deep trough below 1 keV. 
The comparison of the two spectral states shows that the flux transition is to be ascribed entirely to intervening line-of-sight clouds with high column density.

\end{abstract}

\section{Introduction}
\label{sec:intro_335}
 A substantial fraction of Active Galactic Nuclei (AGN) shows absorption from neutral matter in their X-ray spectra (Matt et al. 2002 for a review). If the column density of this gas is higher than 1.5$\times$10$^{24}$~cm$^{-2}$, the level of obscuration is such that below 10~keV the X-ray spectrum is dominated by the reflected radiation and the nucleus is only visible above 10~keV. 
Transitions between Compton-thick (N$_H$$>$1.5$\times$10$^{24}$~cm$^{-2}$) and Compton-thin (N$_H$$<$1.5$\times$10$^{24}$~cm$^{-2}$) states have been reported in some obscured AGNs observed in the X-rays in the recent years (Matt et al. 2003; Guainazzi et al. 2005). The changing look phenomenon may be explained in terms of column density variations, which are frequently observed in type 2 sources (Risaliti et al. 2002) or in terms of the fading of the nuclear emission, as  proposed for NGC4051 (Guainazzi et al. 1998).

The Seyfert 1 galaxy H0557-385 (05$^h$58$^m$02.0$^s$ -38$^\circ$20$^{\prime}$05$^{\prime\prime}$, {\it z}=0.03387) is also known as ESO 055620-3820.2. 
In the X-rays, it was observed by {\it ASCA} (Turner et al. 1996), {\it BeppoSAX} (Quadrelli et al. 2003) and {\it XMM-Newton} (Ashton et al. 2006, hereafter A06). 
In all these observations, the source flux was 1-4$\times$10$^{-11}$~ergs~cm$^{-2}$~s$^{-1}$ and the spectrum appeared to be characterised by a strong warm absorber below 1~keV.  A06 found that a two-phase ionised medium provided a good description of the high resolution data obtained by the Reflection Grating Spectrometer (RGS).

{\it XMM-Newton} re-observed the source twice in 2006, catching it at an extremely low flux state.
This paper presents the analysis of the spectral transition between the {\it XMM-Newton} looks in 2002 and 2006.  	

\section{Observations and data reduction}
\begin{table*}      
\centering                     
\begin{tabular}{c c c c c c c }
\\      
\hline\hline                
 OBSID  & Date          &  pn exp  & $\Gamma$$_{soft}$ &  Flux$_{0.3-2}$  & $\Gamma$$_{hard}$ &  Flux$_{2-10}$ \\    
  -     & (yyyy-mm-dd)  & (ks)     &   - &  (10$^{-12}$ ergs cm$^{-2}$ s$^{-1}$)   & - & (10$^{-12}$ ergs cm$^{-2}$ s$^{-1}$)  \\
\hline\hline 
\\ 
0109130501 & 2002-04-03    &  4    & 1.69$\pm$0.08  &  10.37$\pm$0.07  &  1.60$\pm$0.03 & 36.01$\pm$0.05  \\

0109131001 & 2002-09-17    & 6.5   & 1.53$\pm$0.07   &  10.91$\pm$0.06  &  1.63$\pm$0.02 & 43.30$\pm$0.30 \\

0404260101 & 2006-08-11    &  40   & 2.15$\pm$0.13   & 0.42$\pm$0.20   &  -0.09$\pm$0.07  &  3.06$\pm$0.12 \\                 

0404260301 & 2006-11-03    &  52   &  2.10$\pm$0.15  & 0.37$\pm$0.15 &  -0.02$^{+0.07}_{-0.03}$ &  2.60$\pm$0.15    \\
 \hline                                  
\end{tabular}
\caption{\label{tab:log} Observation Log of H0557-385 as observed by {\it XMM-Newton}. The fluxes have been estimated assuming a soft and a hard X-ray power law model (0.3-2 and 2-10~keV respectively). }
\end{table*}
H0557-385  was observed 4 times (see Table~\ref{tab:log})  by {\it XMM-Newton} (Jansen et al. 2001).
Raw data from all observations were reduced with the Science Analysis System (SAS) using version 8.0. 
The EPIC data of 2002 were collected  in large window mode for the pn camera and in small window mode for the MOS1 and MOS2,
whereas the 2006 EPIC data were all collected in small window.

Background flares were treated according to the method described by Piconcelli et al. 2004 so to maximise 
the signal-to-noise of the data. 
The spectra of the 2002 observations were extracted using pattern 0-4 for the pn and 0-12 for the MOS cameras.
The 0.3-10~keV count rate of 10 counts/s is just below the pile-up threshold in the pn camera.
The source spectra were extracted from a radius of 40 arcsec, yielding a useful exposure 
of 4 and 6.5 ks in the pn instrument. 
No flux variability is found in the light curve and since the spectra from the two 2002 exposures are consistent
in flux and spectral shape, they have been co-added in a single spectral file.

The same procedure for the data reduction has been applied to the 2006 observations.
We note here that due to telemetry loss,  the observation 0404260101 produced two  event lists (scheduled and unscheduled), 
 that were subsequently  merged in a single event list using the SAS task \textsc{merge}.
Source and background spectra and light curves were then extracted from this merged event list.

An inspection of the light curves for (both) the 2006 observations confirms that the source flux  does not vary significantly within 
each of the two exposures, so the analysis will be performed on the  spectra integrated during each epoch.   

As to the RGS data, for the 2002 observations, we will refer to the analysis by A06. 
For the 2006 exposures, RGS data were processed with the task  \textsc{rgsproc} according to the standard method
proposed in the SAS threads.
Due to the flux decrease in 2006, the source was not detected in the RGS images, therefore no high resolution spectra are available for this epoch. 

The source was observed with the Optical Monitor telescope (OM) with optical and UV filters. Photometry is available for the following filters: 5430 (V), 4500 (B), 3440 (U), 2910 (UVW1), 2310 (UVM2), 2120 (UVW2)~$\AA$. 
The data have been analysed with the standard pipeline \textsc{omichain}. The source is detected in all the images in each band.
From the fluxes derived from the observed count-rates, the level of variability is lower than 15-20\% within the two epochs (2002-2006).


\section{Spectral analysis}
In order to apply the $\chi^2$ statistic the EPIC data were rebinned so to oversample the instrumental resolution of a factor $>$3 and to have at least 50 background subtracted counts in each spectral bin for the 2002 data and 25 counts in the 2006 observations.
A Galactic column density of 4$\times$10$^{20}$~cm$^{-2}$ (Dickey \& Lockman, 1990) is included in all the following spectral fits.
Errors are quoted at 90\% (confidence level) for one parameter. Spectral fits are performed over the 0.3-10~keV energy range. 

\begin{figure}
\begin{center}
\psfig{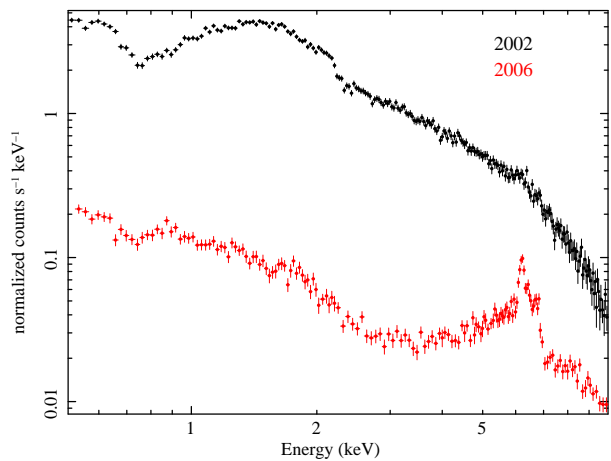}
\caption{\label{fig:two_epochs} EPIC pn spectra of H0557-385 in 2002 and 2006.} 
\end{center} 
\end{figure}

 Figure~\ref{fig:two_epochs} displays the 2002 and 2006 pn data in the full EPIC energy band (0.3-10~keV): unmistakably, the source undergoes  an extraordinary spectral change, with a flux decrease of more than one order of magnitude between 2002 and 2006 (see Table~\ref{tab:log}).
 The two epochs are therefore refereed to as ``high and low flux state" in the following and they are analysed separately.   

The final best fit models for the pn data (which are described in the next sections), have been applied to the MOS spectra in each spectral state. 
This confirms that the MOS spectra are in excellent agreement with the pn data; 
for the sake of brevity,  we quote in the paper only  the parameters obtained from the pn fits.

\subsection{The high flux state: 2002 data}
\label{sec:high_state}
The {\it XMM-Newton} observations of H0557-385 obtained in 2002 were published by A06.
In the analysis presented herein we assume the two warm absorbers model proposed by these authors  based on 
RGS data.
     
The co-added EPIC pn spectrum of the high flux state was fitted  in the 2-10~keV band with a power law continuum and a narrow (zero-width) Gaussian line to account for emission in the Fe K$\alpha$ line at 6.4~keV.  
The best fit parameters are: $\Gamma$=1.65$\pm$0.02, E=6.40$\pm$0.10~keV, line flux of 1.60$\pm$0.70$\times$10$^{-5}$~photons~cm$^{-2}$~s$^{-1}$, with a 2-10~keV luminosity of $\sim$1$\times$10$^{44}$~ergs/s. The Equivalent Width (EW) of the Fe line is 33$^{+21}_{-11}$~eV. 
If the line width is left free to vary  the measured value is $\sigma$=0.20$\pm$0.10~keV, with EW=97$\pm$33~eV, in agreement with the values reported by A06, yielding $\chi$$^2$/d.o.f.=284/225. If the Fe line is fitted with a relativistic profile (\textsc{DISKLINE}, Fabian et al. 1989), the statistic worsens ($\chi$$^2$/d.o.f.=294/226).
The non-zero width of the Fe line may indicate the presence of additional emission components from a higher ionisation state.
We tried to fit the data with two zero-width Gaussians around 6.4~keV and 6.67~keV (mean energy of the Fe~XXV triplet).
The fit statistic is $\chi$$^2$/d.o.f.=290/224, slightly worse than the broad Gaussian fit,   but the  parameters for the Fe XXV line are affected by large uncertainties (E=6.56$\pm$0.62~keV and Flux=1.34$^{+0.56}_{-0.95}$$\times$10$^{-5}$~photons~cm$^{-2}$~s$^{-1}$). The evidence for the presence of a complex profile or of two Fe line components in this spectrum is nonetheless very loose, so we assume the model with a single narrow Gaussian Fe line to compare directly the two spectral states (see Section 4).

When this model is extrapolated  below 2 keV, the data present soft X-ray residuals, therefore  two warm absorbers with ionisation parameters and column densities initially fixed  to the A06 values have been added in the fit.  The model \textsc{zxipcf} available in XSPEC as local model was used. Based  on \textsc{XSTAR}, this model  was originally  developed to describe the effect of a partial covering from ionised matter (Reeves et al. 2008). In the present analysis the covering factor is fixed to 1 so to mimic a  warm  absorber with full coverage. 
Beside the multiphase ionised absorber,  neutral photoelectric absorption  is  required by the data at more than 99.99\% confidence level, consistently to what is reported in A06, so we model it with the XSPEC model \textsc{ZPHABS} obtaining N$_H$=7$\pm$1~$\times$10$^{20}$~cm$^{-2}$.
 The final best fitting parameters are listed in Table~\ref{tab:high_bfit}. 
We briefly note here that  there is a good consistency with the analysis by A06, the small discrepancy  can be probably attributed to the fact that here only EPIC data are considered and to the different models used for fitting the warm absorber.

The spectrum still shows positive residuals around 0.8-0.9~keV; an additional narrow emission line is included in the fit at 0.83$\pm$0.03~keV. The improvement in the fit is $\Delta$$\chi$$^2$=33 for 2 free parameters. 
The reader is deferred to Section~\ref{sec:disc} for the discussion on the identification of this feature.

\begin{table}      
\begin{center}    
\caption{\label{tab:high_bfit} Best fit parameters of the high flux state (pn data) fitted with the following model: 
{\small POWER LAW + NEUTRAL N$_H$ +2 WARM ABSORBERS (ZXIPCF) + 2 GAUSSIANS}}                 
\begin{tabular}{c c}    
\hline\hline    
2002 spectrum  &  $\chi^2$/d.o.f.=294/226 \\    
\hline
\\               
$\Gamma$=1.87$\pm$0.01 & N$_H$ (neutral)=0.07$\pm$0.01   \\

N$_H$(warm$_1$)=1.02$^{+0.10}_{-0.30}$  & Log$\xi$$_1$=2.16$\pm$0.06 \\

N$_H$(warm$_2$)=0.60$\pm$0.03  & Log$\xi$$_2$=0.56$\pm$0.05 \\

E$_1$=6.40$\pm$0.11  & Flux$_1$=1.51$\pm$0.71       \\

 E$_2$=0.83$\pm$0.02  & Flux$_2$=4.50$\pm$1.02   \\     
\hline 
\end{tabular}
\end{center}
Line fluxes in units of $\times$10$^{-5}$ ph~cm$^{-2}$~s$^{-1}$; Line energies in keV; Column densities in $\times$10$^{22}$~cm$^{-2}$; Ionisation parameter in ergs~cm~s$^{-1}$. \\
\end{table} 

\subsection{The low flux state: 2006 data}
\label{sec:low_state}
As reported in Table~\ref{tab:log}, the two observations of 2006 were performed with a 3 months gap in between, making it necessary to check for any spectral variation between the two spectra. 
In the following, we will refer to the 2006 spectra as Aug06 and Nov06. 
At first glance, the two spectra are very similar in shape and flux intensity. 

The spectral shape in Figure~\ref{fig:two_epochs} suggests that the drop of counts from the high state could be due to absorption in the 1-6~keV band.
We therefore start by analysing the longest spectrum (Nov06) and by constructing a model where the 2002 best fit is absorbed by neutral gas. The addition of a cold column density to account for the absorption (model  \textsc{ZPHABS}, already described in the previous section) does not provide a good description of the data ($\chi^2$/d.o.f.=896/233).   
It seems that the nuclear primary continuum is not totally suppressed, but it still contributes to the soft X-ray spectrum,
therefore a partial covering absorber is considered. The partial covering is constructed by adding a power law convolved with the \textsc{ZPHABS} and \textsc{CABS} components in XSPEC. The latter is necessary to produce a correct estimate of the continuum normalisation by taking into account Compton scattering by optically thin gas. 
This partial covering model is able to reproduce the spectral shape correctly, with  column density of 8$\pm$0.2$\times$10$^{23}$~cm$^{-2}$ and covering fraction $>$ 93 percent (see Table~\ref{tab:low_bfit}).

\begin{figure}
\begin{center}
\psfig{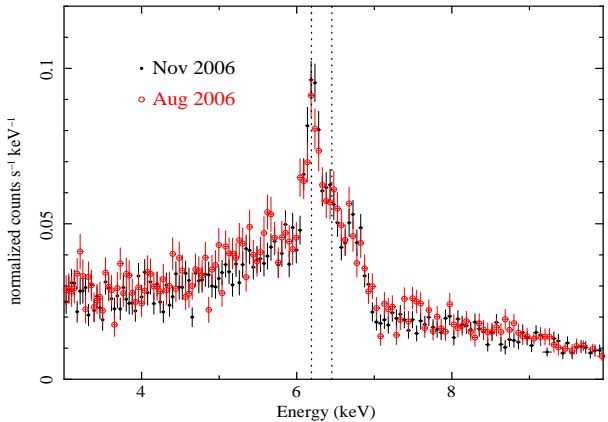}
\caption{\label{fig:k_alpha}EPIC pn spectra of the low flux state in the Fe~K band. The dotted lines mark the position of the K$\alpha$  transitions for Fe I and Fe XXV (6.4 and 6.67~keV).} 
\end{center} 
\end{figure}

The remarkable transition to a low flux state  uncovers a prominent  Iron emission line complex  (see Figure~\ref{fig:k_alpha}), suggesting that Compton reflection could also contribute to the hard X-ray emission.  
 We therefore added a pure Compton reflection \textsc{PEXRAV} component (Magdziarz \& Zdziarski 1995) assuming solar abundances and a 30$^{\circ}$ inclination of the reflecting material to the observer line of sight.
This component is very significant and it improves the fit statistic by $\Delta$$\chi$$^2$=268 for 1 free parameter.
The reflection fraction (i.e. R=$\Omega$/2$\pi$)  estimated as the ratio of the \textsc{PEXRAV} and the direct power law normalisations, is R=0.57$\pm$0.04.

Four zero-width Gaussian lines have been included to fit the Fe~K complex: K$\alpha$ and K$\beta$ transitions from neutral Fe at 6.4 and 7.06~keV, and K$\alpha$ lines from Fe~XXV-XXVI at 6.67  and 6.96~keV. The parameters for the detected Fe lines are reported in Table~\ref{tab:low_bfit}.
The Fe~XXVI  K$\alpha$ and  Fe~I K$\beta$ are not detected in the spectra, so they are not included in the table. The upper limit on their fluxes were obtained by fixing the peak energy to its laboratory value and they are respectively 2.2$\times$10$^{-6}$ and 2.5$\times$10$^{-6}$~ph~cm$^{-2}$s$^{-1}$.

The neutral absorber observed in the high state with N$_H$$\sim$7$\times$10$^{20}$~cm$^{-2}$, is not detected  in the 2006 data, with an upper limit on the column density  of $\sim$3$\times$10$^{19}$~cm$^{-2}$.

The effect of the soft X-ray absorbers is still very evident in the data, so the same warm absorbers used for the high state are included in the model. 
An additional Gaussian line is required to fit residuals around 0.9~keV and it has been included in the fit ($\Delta$$\chi$$^2$=83 for 2 free parameters). 

This  model provides a satisfactory fit for  Nov06 and it was  then applied  to Aug06.
The best fitting parameters for both spectra are reported in Table~\ref{tab:low_bfit}.
The two spectra are in very good agreement, therefore we can safely assume that the source was not affected by significant spectral change between August and November 2006. 

  The drastic change in the X-ray emission could also be explained if the primary emission simply drops, letting only the reflected spectrum be observable (Matt et al. 2003). This hypothesis is tested by fitting the November data with a PEXRAV component  to account for the hard X-ray spectrum,  and a single power law to model the soft X-ray excess. The warm absorber components and all the Gaussian lines reported in Table~\ref{tab:low_bfit} are included in the fit. This model yields an acceptable fit statistic ($\chi^2$/d.o.f.=274/233), but the intrinsic photon index of the reflected component is unrealistic ($\Gamma$=1.13$\pm$0.03). If the PEXRAV photon index is fixed to the high state intrinsic power law ($\Gamma$=1.9), the $\chi^2$ worsens considerably, being
$\chi^2$/d.o.f.=458/234. These results allow us to exclude that H0557-385 was reflection dominated in the 2006 low state, therefore this interpretation will not be discussed in the remainder of the paper.

\begin{table}      
\begin{center}    
\caption{\label{tab:low_bfit} Best fit parameters of the low flux state  (pn data) fitted with the following model: 
{\small POWER LAW +2 WARM ABSORBERS (ZXIPCF) + PEXRAV + PARTIAL COVERING (CABS*ZPHABS*PLAW) + 3 GAUSSIANS}}                 
\begin{tabular}{c c}    
\hline\hline                
  AUGUST 2006  &  $\chi^2$/d.o.f.=249/217 \\    
\hline
\\ 
$\Gamma$=1.94$^{+0.10}_{-0.19}$ & R=0.73$\pm$0.01  \\

N$_H(pc)$=81$\pm$2 & C$_{f}$$>$0.93 \\ 

N$_H$(warm$_1$)=1.02$^{+0.85}_{-0.40}$  & Log$\xi$$_1$=2.28$^{+0.24}_{-0.16}$ \\

N$_H$(warm$_2$)=0.10$^{+0.09}_{-0.02}$  & Log$\xi$$_2$=0.28$^{+0.42}_{-0.32}$ \\

E$_1$=6.43$^{+0.01}_{-0.14}$  &  Flux$_1$=1.08$\pm$0.17      \\

 E$_2$=6.66$\pm$0.06  & Flux$_2$=0.26$\pm$0.15  \\
 
E$_3$=0.89$\pm$0.01  & Flux$_3$=1.52$\pm$0.36 \\  
     
\hline\hline

NOVEMBER 2006  & $\chi^2$/d.o.f.=263/232  \\
\hline
\\
$\Gamma$=1.95$^{+0.02}_{-0.19}$ & R=0.57$\pm$0.04   \\

N$_H(pc)$=80$\pm$2 &  C$_{f}$$>$0.93 \\ 

N$_H$(warm$_1$)=1.25$^{+0.14}_{-0.08}$  & Log$\xi$$_1$=2.15$\pm$0.07 \\

N$_H$(warm$_2$)=0.10$\pm$0.01  & Log$\xi$$_2$=0.28$^{+0.08}_{-0.04}$ \\

E$_1$=6.41$^{+0.02}_{-0.01}$  &  Flux$_1$=1.40$\pm$0.15      \\

 E$_2$=6.62$\pm$0.03  & Flux$_2$=0.42$\pm$0.13  \\
 
E$_3$=0.90$\pm$0.01  & Flux$_3$=1.60$\pm$0.28 \\

\hline\hline  
\end{tabular}
\end{center}
Line fluxes in units of $\times$10$^{-5}$ ph~cm$^{-2}$~s$^{-1}$; Line energies in keV; Column densities in $\times$10$^{22}$~cm$^{-2}$; Ionisation parameter in ergs~cm~s$^{-1}$. \\
\end{table}

\section{Discussion}
\label{sec:disc}
The spectral analysis in the previous section reveals a peculiar behaviour for this Seyfert~1 
galaxy. From a general overview of Tables~\ref{tab:high_bfit} and \ref{tab:low_bfit}, it is clear that 
most of the fitting parameters do not vary between the 2002 and the 2006 spectra. 
The photon index of the power law and the warm absorbers column density and ionisation state 
are remarkably consistent within the two epochs. 
Notably, the 2002 data require a simpler model (Table~\ref{tab:high_bfit}) compared to the one 
found for the 2006 data in Table~\ref{tab:low_bfit}. 
Nonetheless, for a more direct comparison of the two states, the low state model is applied to the high state spectrum, as it follows.

The normalisation of the power law at 1~keV is around 1.45$\pm$0.05$\times$10$^{-2}$~ph~cm$^{-2}$~s$^{-1}$ and 1.02$\pm$0.05$\times$10$^{-2}$~ph~cm$^{-2}$~s$^{-1}$
 for the high and low state, respectively.
These numbers imply that the intrinsic X-ray photons flux remains visible in both states with a variation lower than 50 percent between 2002 and 2006. 
As described in Section~\ref{sec:low_state}, a fraction higher than 93 percent  of the intrinsic flux is absorbed by circumnuclear material with high column density ($\sim$~8$\times$10$^{23}$~cm$^{-2}$) during the low state.
 The high state data do not show evidence for partially covering gas.  When applying  the low state model to these data and after fixing the column density to 8$\times$10$^{23}$~cm$^{-2}$ (i.e. the low state value),  the upper limit on the covering fraction   is about 5~percent.
The emerging picture is that of an active nucleus  covered by intervening optically thick cold gas in 2006, which 
 may have been present during the high state epoch with a lower column and/or with sparse coverage, as tested by applying the partial covering model to the 2002 data. 

 The flux of the  narrow Fe~K$\alpha$ line at high and low state  are in good agreement (Tables~\ref{tab:high_bfit} and \ref{tab:low_bfit}).
The absence of a Fe~K$\alpha$ broad relativistic component and the lack of Fe K variability seem to point to a scenario in which the Compton reflection occurs on the parsec-scale torus-shaped material expected to be present in Seyfert galaxies (Antonucci 1993).     
The Fe line intensity measured with respect to the reflection component is  EW=460$^{+378}_{-215}$~eV in the high state and 
EW=510$\pm$55~eV in the low state.
Both values are considerably lower than the theoretical 1~keV Fe line expected from reflection in a  Compton-thick torus
(Ghisellini et al. 1994). Alternatively, a substantial fraction of the Fe line could be emitted by transmission in the obscuring clouds. Anyhow, it is very difficult to separate Fe line emission  from the torus and from the clouds without a clear picture of the AGN geometry.

We can speculate on the distance of the clouds by estimating  their velocity from the FWHM of the Fe K line at low state. The upper limit on the line width is 45~eV, resulting in a velocity of 5000~km~s$^{-1}$.  The source is known to have several components of the H$\alpha$ line (A06 and reference therein) with FWHM of 1035, 2772 and 11000~km~s$^{-1}$, suggesting a location of the obscuring clouds consistent with that of the two lower velocity components of the Broad Line Region. Assuming that in 2006 the source remained at low state for 3 months, this gives a minimum duration of the occultation event.  The clouds can then be located at a  distance higher than {\it v$\times$t}=5000~km~s$^{-1}$$\times$3 months i.e. $\sim$3.9$\times$10$^{15}$~cm. 

Determining more detailed properties of the system such as the precise time scale for the transit of the cloud or a characteristic size, appears very difficult since H0557-385 was observed very sparsely 
 by X-ray observatories. The closest look to the  {\it XMM-Newton} ones, is an ASCA observation
 of 1995 (Turner et al. 1996) which shows the source at high state.
 In the attempt to find additional X-ray data, H0557-385 has been searched for in the {\it XMM-Newton} Slew Survey images (Saxton et al. 2008); luckily, one recent slew (25/02/2008)  passed through  the coordinates  corresponding to the  AGN (Saxton, private communication). H0557-385 was not detected, but a reliable  upper limit  on the flux could be derived: 
 $<$1.5$\times$10$^{-12}$~ergs~cm$^{-2}$~s$^{-1}$ for the 0.2-2~keV band and $<$1$\times$10$^{-11}$~ergs~cm$^{-2}$~s$^{-1}$ in the 2-12~keV band.   Arguably, this demonstrates that after $\sim$2 years, the source was still at low state. Whether it has remained in this state since 2006 through February 2008 is, obviously, impossible to tell.

  From the analysis of the UV data (Section 2) it is found that the source of UV photons remains essentially unaffected by the X-ray change.  This implies that the size of the X-ray source must be very compact compared  to that of the UV source.

Highly obscured AGN constitute an excellent laboratory to study the physical 
properties of the circumnuclear gas via soft X-ray spectroscopy of emission lines produced in the interaction between such gas and the primary nuclear emission (Guainazzi \& Bianchi 2007).
Unfortunately, the high resolution cameras did not provide a detection for  H0557-385 at low state  and  so the source does not have the wealth of additional spectral information that was available for other Seyfert~1 caught at low state, e.g. Mrk~335 (Longinotti et al. 2008), NGC~4051 (Pounds et al. 2004).
Nonetheless, the 2006 EPIC data do show an emission line around 0.9~keV, which is consistent with emission from  the NeIX triplet.
We have searched for this component in the RGS spectrum of 2002. Assuming the best fit model in Table~\ref{tab:high_bfit}, the line is detected with Flux=3.41$^{+1.92}_{-2.34}$$\times$10$^{-5}$~ph~cm$^{-2}$~s$^{-1}$ and  E=0.916$\pm$0.017~keV.

In addition to the NeIX line, the 6.6~keV emission line identified as Fe XXV in the low state spectrum could also be a signature of photoionisation. However,  the statistical quality of these measurements prevent us from drawing any firm conclusion on the origin of these spectral features. 

 Detecting soft X-ray emission (both lines and continuum) may lead to interpret H0557-385 at low state within the typical Seyfert 2 scenario (Bianchi et al. 2006; Guainazzi \& Bianchi 2007).  Remarkably, the presence of the same two phase-warm absorber in {\it both} states  plays a key role in ruling out this scenario.  It provides a very robust argument  against the "Seyfert 2/scattering" scenario and consequently, in favour of the partial covering. The fact that the warm gas is still visible through the absorption at low state, implies that the soft X-ray emission does not come from an extended region as the NLR in obscured AGN (it would be very unlikely to have a warm absorber at such large scale). Accordingly, the soft X-ray emission must have been only partially reduced by the obscuring event, as also suggested by the comparison of the power law normalisations, and therefore it is still observable at low state. 

The fact that the Seyfert~1 nucleus remains visible at low state makes the spectral transition in H0557-385 a very interesting event, which resembles the occultation phenomenon seen in the Seyfert~2 galaxy NGC1365 (Risaliti et al. 2005, 2007), but it is the presence of the warm absorber what makes H0557-385 a unique case where the partial covering scenario is indeed very robust and can be challenged only if other  spectral states were observed.


\section*{Acknowledgments}
This paper is based on observations obtained with {\it XMM-Newton}, an ESA science mission with instruments and contributions directly funded by ESA Member States and NASA. We would like to thank the referee for his/her valuable and careful  comments. We thank Pedro Rodr\'iguez-Pascual for his help in handling the OM data. 
We acknowledge support from the Faculty of the European Space Astronomy Centre (ESAC).
SB acknowledges financial support from ASI (grant I/088/06/0).

\end{document}